\documentclass[12pt]{article}
\usepackage{amsfonts}
\usepackage{amssymb,amsmath,amsthm,latexsym}
\usepackage[numbers,compress]{natbib}
\usepackage{amsmath}
\usepackage{indentfirst}
\allowdisplaybreaks[4]
\newtheorem{theorem}{Theorem}[section]

\newtheorem{lemma}[theorem]{Lemma}

\newtheorem{define}[theorem]{Definition}
\newtheorem{example}[theorem]{Example}

\newcommand{\EOP} { \hfill $\Box$ }

\newcommand{\pf} { {\rm \noindent{\bf Proof.}} }

\numberwithin{equation}{section}

\begin{document}
\title{ The weight distribution of a family of  $p$-ary cyclic codes   }
\author{Long Yu, Hongwei Liu{\thanks{Corresponding author. \newline  Email address:~hwliu@mail.ccnu.edu.cn~(Hongwei Liu),~longyu@mails.ccnu.edu.cn~(Long Yu).} }}
\date{School of Mathematics and Statistics, Central China Normal University, Wuhan, Hubei 430079, China}
\maketitle
\begin{abstract}
Let $m$, $k$ be  positive integers, $p$ be an odd prime and $\pi $ be a primitive element
of $\mathbb{F}_{p^m}$. In this paper, we determine the weight distribution of a family of  cyclic codes $\mathcal{C}_t$ over $\mathbb{F}_p$, whose duals have two zeros $\pi^{-t}$ and $-\pi^{-t}$, where $t$ satisfies $t\equiv \frac{p^k+1}{2}p^\tau \ ({\rm mod}\ \frac{p^m-1}{2})  $ for some  $\tau \in \{0,1,\cdots, m-1\}$.
\end{abstract}


\section{Introduction}
Let $p$ be an odd prime and $q$ be a power of $p$. An $[n,k,d]$ linear code over the finite field $\mathbb{F}_q$ is a $k$-dimensional subspace of $\mathbb{F}_{q}^{n}$ with minimum Hamming distance $d$.  A linear code $\mathcal{C}$ of length $n$ is called cyclic if $(c_0,c_1,\cdots,c_{n-1})\in \mathcal{C}$ implies that $ (c_{n-1},c_0,\cdots,c_{n-2}) \in \mathcal{C}$. By identifying a codeword $(c_0,c_1,\cdots,c_{n-1})\in \mathcal{C}$ with the polynomial
\[c_0+c_1X+\cdots+c_{n-1}X^{n-1}\in \mathbb{F}_q[X]/(X^n-1),\]
a cyclic code $\mathcal{C}$ of length $n$ over $\mathbb{F}_q$ corresponds to an ideal of $\mathbb{F}_q[X]/(X^n-1)$. The monic generator $g(X)$ of this ideal is called the generator polynomial of $\mathcal{C}$, which satisfies that $g(X)|(X^n-1)$. The polynomial $h(X)=(X^n-1)/g(X)$ is referred to as the parity-check polynomial of $\mathcal{C}$ \cite{macwilliams}.

Let $A_i$ denote the number of codewords with Hamming weight $i$ in a linear code $\mathcal{C}$ of length $n$. The weight enumerator of $\mathcal{C}$ is defined by
\[A_0+A_1X+A_2X^2+\cdots+A_{n}X^{n}, \ \ \ \ {\rm where}\ A_0=1.\]
The sequence $(A_0,A_1,\cdots,A_n)$ is called the weight distribution of the code $\mathcal{C}$. In general the weight
distribution of cyclic codes are difficult to be determined and they are known only for a few special classes. There are some results on  the weight distribution of
cyclic codes whose duals have two or more zeros( see \cite{Ding2011,Ding2013,FengandLuo,Feng,Lichulei,LuoandFeng1,LuoandFeng2,Rao,Vega12,Vega07,Wang,XiongDCC,Xiong2013,Xiong2012,Zeng2012,Zeng,zheng,zhou,zhouIT}).

Let $\mathbb{Z}_m$ be the residue ring modulo integer $m$,  $\Gamma_i$ be the $p$-cyclotomic coset modulo $p^m-1$ containing $i$, i.e.,
 \[ \Gamma_i=\{i\cdot p^j\ ({\rm mod}\  p^m-1) \mid j=0,1,\cdots,m-1 \},\]
where $i \in \mathbb{Z}_{p^m-1}$. A subset $\{i_1, i_2, \cdots, i_r\}$ of $\{0,1,\cdots,p^m-2\}$, where
$r\geq 1$, is called a \emph{complete set of representatives} of all $p$-cyclotomic cosets modulo $p^m-1$ if $\Gamma_{i_1},\Gamma_{i_2},\cdots, \Gamma_{i_r}$ are pairwise disjoin and $\bigcup_{k=1}^r\Gamma_{i_k}=\{0,1,\cdots, p^m-2\}$.

Throughout this paper,  we assume that $m$, $k$ are  positive integers, $\pi$ is a primitive element of $\mathbb{F}_{p^m}$ and $\zeta_p$ is a primitive  $p$-th unity root.
Let $t$ be an integer such that $(\pi^{t})^{p^i}\neq -\pi^{t} $ for all $i\in \mathbb{Z}_m$.
Let $h_1(x)$ and $h_2(x)$ be the minimal polynomials of $\pi^{-t}$ and $-\pi^{-t}$
over $\mathbb{F}_p$, respectively.
In this paper, we let $h(x)=h_1(x)h_2(x)$ and  $\mathcal{C}_t$
be the cyclic code with parity-check polynomial $h(x)$.
By the well-known Delsarte's Theorem~\cite{Delsarte}, cyclic code $\mathcal{C}_t$ can be expressed as
\begin{equation}\label{eq:cycliccode}
\mathcal{C}_t=\left\{\textbf{c}(a,b)=\left({\rm Tr}_1^m(a\pi^{ti}+b(-\pi^t)^{i})\right)_{i=0}^{p^m-2}\mid a,b \in \mathbb{F}_{p^m}\right\},
\end{equation}
where ${\rm Tr}_1^m(\cdot)$ is the trace function from $\mathbb{F}_{p^m}$ to $\mathbb{F}_{p}$. In particular, in the case of $t=1$, Vega and  Wolfmann in \cite{Vega07} have studied cyclic code  $ \mathcal{C}_1$. They constructed this class of   cyclic codes   from the direct sum of two one-weight irreducible cyclic codes for odd $m$ and proved that $\mathcal{C}_1$ has only two nonzero weights. In \cite{Ma}, Ma et al. further investigated cyclic code $ \mathcal{C}_1$ for the case of even $m$, and gave the weight distribution of the code $ \mathcal{C}_1$.

The goal of this paper is to determine the weight distribution of a family of cyclic codes $\mathcal{C}_t$ over $\mathbb{F}_p$ defined by (\ref{eq:cycliccode}) in the case of $t\equiv \frac{p^k+1}{2} p^{\tau} \  ({\rm mod}\ \frac{p^m-1}{2})$ for some $\tau \in \mathbb{Z}_m$.  By applying    the value distribution of the exponential sum $\sum\limits_{x\in \mathbb{F}_{p^m}}\zeta_p^{{\rm Tr}_1^m(\alpha x^{p^k+1})},\ \alpha\in \mathbb{F}_{p^m}$ , which is given in \cite{Draper,Helleseth-Kholosha2006,Lidl R}, we obtain the value distribution of the exponential sum $\sum\limits_{u\in \mathbb{F}_p^*}\sum\limits_{x\in \mathbb{F}_{p^m}}(\zeta_p^{u{\rm Tr}_1^m((a+b) x^{p^k+1})}+$ $\zeta_p^{u{\rm Tr}_1^m((a-b) x^{p^k+1})}),\ a,\ b \in\mathbb{F}_{p^m}$. Based on these results,  we characterize the weight distribution of the code $\mathcal{C}_t$ defined by (\ref{eq:cycliccode}).

This paper is organized as follows. Section $\textbf{2}$ gives some basic definitions and  results  over finite fields. In Section $\textbf{3}$,  we determine the weight distribution of a family of cyclic codes $\mathcal{C}_t$ defined by~(\ref{eq:cycliccode})  over $\mathbb{F}_p$.

\section{Preliminaries }
Let $\mathbb{F}_q$ be a finite field and $\mathbb{F}_q^*=\mathbb{F}_q \setminus \{0\}$, where $q$ is a power of an odd prime $p$. In the following, we give a brief introduction to the theory of quadratic forms over finite fields, which is needed to calculate the weight distribution of the cyclic codes in
the next section. Quadratic forms have been well studied (\cite{Lidl R}), and they have many applications in sequence design  and coding theory.
\begin{define}
Let $\{\omega_1, \omega_2, \cdots, \omega_s\}$ be a basis for $\mathbb{F}_{q^s}$
over $\mathbb{F}_q$ and $x=\sum_{i=1}^s x_i\omega_i$, where $x_i \in \mathbb{F}_q$. A function $f(x)$ from $\mathbb{F}_{q^s}$
to $\mathbb{F}_q$ is called  quadratic form if it can be
represented as $$f(x)=f(\sum_{i=1}^s x_i\omega_i)=\sum_{1\leq i\leq j\leq s}a_{ij}x_ix_j ,\ \ \ \ a_{ij}\in \mathbb{F}_q.$$
\end{define}
\noindent The rank of the quadratic form $f(x)$ is defined as the codimension of
the $\mathbb{F}_q$-vector space
\[ V=\{ x\in \mathbb{F}_{q^s} \mid f(x+z)-f(x)-f(z)=0,\ \ {\rm for \ \ all }\ z\in \mathbb{F}_{q^s}\}, \] denotes by rank $(f)$. Then $|V|=q^{s-{\rm rank}(f)}$.

For a quadratic form $f (x)$ with $s$ variables over $\mathbb{F}_q$, there exists a symmetric matrix $A$ of order $s$ over $\mathbb{F}_q$ such that $f(x)=XAX'$, where $X=(x_1,x_2,\cdots,x_s)\in \mathbb{F}_q^s$ and $X'$
denotes the transpose of $X$. It is known that there exists a nonsingular matrix $T$ over $\mathbb{F}_q$ such that $TAT'$ is a diagonal matrix \cite{Lidl R}. Making a nonsingular linear substitution $X=ZT$ with $Z=(z_1,z_2,\cdots,z_s)\in \mathbb{F}_q^s$, we have
\[f(x)=Z(TAT')Z'=\sum_{i=1}^ra_iz_i^2,\ \ \ \ a_i\in \mathbb{F}_q^*,\]
where $r$ is the rank of $f(x)$.
%

Let  $v_2(j)$ denote the  $2$-adic valuation of integer $j$ (i.e., the maximal power of $2$ dividing $j$). We denote the quadratic character of $\mathbb{F}_{p^m}$ by \[\eta_m(x)=\left\{\begin{array}{ll}
                          0, & \hbox{if $x=0$;} \\
                          x^{\frac{p^m-1}{2}}, & \hbox{if $x\in\mathbb{F}_{p^m}^*$.}
                        \end{array}
                      \right.\]
\begin{lemma}\label{lem:gcdkm}{\rm \cite{Draper,Helleseth-Kholosha2006}}
Let $k$, $m$ be positive integers. Then
\[\gcd(p^k+1,p^m-1)=\left\{
  \begin{array}{ll}
    p^{\gcd(k,m)}+1, & \hbox{if $v_2(m)>v_2(k)$;} \\
    2, & \hbox{otherwise.}
  \end{array}
\right.\]
\end{lemma}
The following lemma could be found in \cite{Draper,Helleseth-Kholosha2006,Lidl R}.
\begin{lemma}\label{lemTfenbu}{\rm \cite{Draper,Helleseth-Kholosha2006,Lidl R}}
Let $T_\alpha(x)=\sum\limits_{x\in \mathbb{F}_{p^m}}\zeta_p^{{\rm Tr}_1^m(\alpha x^{p^k+1})} $, where $k$ is a positive integer.
Then for any $\alpha\in \mathbb{F}_{p^m}$,
\begin{itemize}
\item if $v_2(k)\geq v_2(m)$, then
\[T_\alpha(x)=\left\{
  \begin{array}{ll}
    \eta_m(\alpha)(-1)^{m-1}p^{\frac{m}{2}}, & \hbox{if $p\equiv 1 \ ({\rm mod}\ 4)$;} \\ \\
    \eta_m(\alpha)(\sqrt{-1})^{m}(-1)^{m-1}p^{\frac{m}{2}}, & \hbox{if $p\equiv 3 \ ({\rm mod}\ 4)$.}
  \end{array}
\right.\]
\item if $v_2(k)+1=v_2(m)$, then
  \[T_\alpha(x)=\left\{
    \begin{array}{ll}
      p^{\frac{m}{2}+d}, & \hbox{$\frac{p^m-1}{p^d+1}$ times;} \\
      -p^{\frac{m}{2}}, & \hbox{$\frac{p^d(p^m-1)}{p^d+1}$ times;}\\
      p^m, & \hbox{$1$ time.}
    \end{array}
  \right.\]
  \item if $v_2(k)+1<v_2(m)$, then
  \[T_\alpha(x)=\left\{
    \begin{array}{ll}
      -p^{\frac{m}{2}+d}, & \hbox{$\frac{p^m-1}{p^d+1}$ times;} \\
      p^{\frac{m}{2}}, & \hbox{$\frac{p^d(p^m-1)}{p^d+1}$ times;}\\
      p^m, & \hbox{$1$ time.}
    \end{array}
  \right.\]
\end{itemize}
\end{lemma}

\section{The weight distribution of the code $\mathcal{C}_t$}
In this section, we let $n=p^m-1$. For a given positive divisor $l$ of $m$, the
trace function from $\mathbb{F}_{p^m}$ to $\mathbb{F}_{p^l}$ is defined by ${\rm Tr}_l^m(x)=\sum_{i=0}^{\frac{m}{l}-1}x^{p^{li}}$, where $x\in \mathbb{F}_{p^m}$.
Let $SQ$ denote the set of  square elements in $\mathbb{F}_{p^m}^*$,  $SQ_p$ denote the set of square elements in $\mathbb{F}_{p}^*$ and let  $u_p$ be a primitive element in $\mathbb{F}_p$.
%
%
In the following, we compute the weight of the codeword $\textbf{c}(a,b)\in \mathcal{C}_t$ defined by (\ref{eq:cycliccode}).
\begin{eqnarray}\label{eq:wtcab}
\nonumber  wt(\textbf{c}(a,b))&= & \# \{0\leq i\leq p^m-2: c_i \neq0\} \\
\nonumber   &=& n-\frac{1}{p}\sum_{i=0}^{p^m-2}\sum_{u\in \mathbb{F}_p}\zeta_p^{u {\rm Tr}_1^m(a\pi^{ti}+b(-\pi^t)^{i})} \\
\nonumber   &=& n-\frac{1}{p}\sum_{u\in \mathbb{F}_p}\sum_{i=0}^{\frac{p^m-3}{2}}\left(\zeta_p^{u {\rm Tr}_1^m(a\pi^{2ti}+b(-\pi^t)^{2i})}+\zeta_p^{u {\rm Tr}_1^m(a\pi^{(2i+1)t}+b(-\pi^t)^{(2i+1)})}\right) \\
\nonumber   &=& n-\frac{1}{p}\sum_{u\in \mathbb{F}_p}\sum_{x\in SQ}\left(\zeta_p^{u {\rm Tr}_1^m(ax^{t}+bx^{t})}+\zeta_p^{u {\rm Tr}_1^m(a(\pi x)^{t}-b(\pi x)^{t})}\right) \\
\nonumber   &=& n-\frac{1}{2p}\sum_{u\in \mathbb{F}_p}\sum_{x\in \mathbb{F}_{p^m}^*}\left(\zeta_p^{u {\rm Tr}_1^m(ax^{2t}+bx^{2t})}+\zeta_p^{u {\rm Tr}_1^m(a\pi^{t} x^{2t}-b\pi^{t} x^{2t})}\right) \\
 &=&p^m-p^{m-1}-\frac{1}{2p}\sum_{u\in \mathbb{F}_p^*}\sum_{x\in \mathbb{F}_{p^m}}\left(\zeta_p^{u {\rm Tr}_1^m((a+b)x^{2t})}+\zeta_p^{u {\rm Tr}_1^m((a-b)\pi^{t} x^{2t})}\right).
\end{eqnarray}

Therefore, we have the following lemma.
\begin{lemma}\label{lem:zongfenlei}
With the notations given above. If $t, e \in \mathbb{Z}_{p^m-1}$ are two positive integers such that $t\equiv e p^\tau \  ({\rm mod}\ \frac{p^m-1}{2})$ for some $\tau\in \mathbb{Z}_m$, then the cyclic codes $\mathcal{C}_t$ and $\mathcal{C}_e $ defined by (\ref{eq:cycliccode}) have the same weight distribution.
\end{lemma}
\pf By (\ref{eq:wtcab}), we have that the weight distribution of $\mathcal{C}_t$ and $\mathcal{C}_e$ are respectively determined by the value
distribution of
\[\Delta_t= \sum_{u\in \mathbb{F}_p^*}\sum_{x\in \mathbb{F}_{p^m}}\left(\zeta_p^{u {\rm Tr}_1^m((a+b)x^{2t})}+\zeta_p^{u {\rm Tr}_1^m((a-b)\pi^{t} x^{2t})}\right)\]
and \[\Delta_e=\sum_{u\in \mathbb{F}_p^*}\sum_{x\in \mathbb{F}_{p^m}}\left(\zeta_p^{u {\rm Tr}_1^m((a+b)x^{2e})}+\zeta_p^{u {\rm Tr}_1^m((a-b)\pi^{e} x^{2e})}\right).\]
Let $e\equiv r p^{m-\tau} \  ({\rm mod}\ p^m-1)$, then  the integers $t$ and $r$ satisfy $t\equiv r \  ({\rm mod}\ \frac{p^m-1}{2}) $, i.e. $t=r+l\frac{p^m-1}{2} $ for some integer $l$. Hence
\begin{eqnarray*}
  \Delta_t &=&  \sum_{u\in \mathbb{F}_p^*}\sum_{x\in \mathbb{F}_{p^m}}\left(\zeta_p^{u {\rm Tr}_1^m((a+b)x^{2t})}+\zeta_p^{u {\rm Tr}_1^m((a-b)\pi^{t} x^{2t})}\right)\\
   &=&  \sum_{u\in \mathbb{F}_p^*}\sum_{x\in \mathbb{F}_{p^m}}\left(\zeta_p^{u {\rm Tr}_1^m((a+b)x^{2r+l(p^m-1)})}+\zeta_p^{u {\rm Tr}_1^m((a-b)\pi^{r+l\frac{p^m-1}{2}} x^{2r+l(p^m-1)})}\right)\\
   &=& \sum_{u\in \mathbb{F}_p^*}\sum_{x\in \mathbb{F}_{p^m}}\left(\zeta_p^{u {\rm Tr}_1^m((a+b)x^{2r})}+\zeta_p^{u \pi^{l\frac{p^m-1}{2}}{\rm Tr}_1^m((a-b) \pi^{r} x^{2r})}\right) \\
   &=&  \sum_{u\in \mathbb{F}_p^*}\sum_{x\in \mathbb{F}_{p^m}}\left(\zeta_p^{u {\rm Tr}_1^m((a+b)x^{2r})}+\zeta_p^{u {\rm Tr}_1^m((a-b)\pi^{r} x^{2r})}\right)\\
   &=&  \Delta_r
\end{eqnarray*}
where the fourth identity is obtained by $u\pi^{l\frac{p^m-1}{2}}=\pm u$.
As we know, $\{\Delta_r\mid a,b \in \mathbb{F}_{p^m}\}=\{\Delta_e\mid a,b \in \mathbb{F}_{p^m}\}$, since $e\equiv r p^{m-\tau} \  ({\rm mod}\ p^m-1)$. Therefore, the multi-sets $\{\Delta_t\mid a,b \in \mathbb{F}_{p^m}\}$ and $\{\Delta_e\mid a,b \in \mathbb{F}_{p^m}\}$ have the same value distribution.\EOP

In this section, we study  the weight distribution of  cyclic codes $\mathcal{C}_t$ in the case of $t\equiv \frac{p^k+1}{2} p^{\tau} \  ({\rm mod}\ \frac{p^m-1}{2})$ for some $\tau\in \mathbb{Z}_m$. By Lemma~\ref{lem:zongfenlei}, we have that the codes $ \mathcal{C}_t$ and $ \mathcal{C}_{\frac{p^k+1}{2}}$ have the same weight distribution. In order to determine the weight distribution of the code  $ \mathcal{C}_t$, we just need to obtain the weight distribution of the code  $ \mathcal{C}_{\frac{p^k+1}{2}}$. In the following, we calculate the  weight distribution of  cyclic code $\mathcal{C}_{\frac{p^k+1}{2}}$.
By (\ref{eq:cycliccode}), we have
\begin{equation}\label{eq:Cpk+1}
\mathcal{C}_{\frac{p^k+1}{2}}=\left\{\textbf{c}(a,b)=\left({\rm Tr}_1^m(a\pi^{i\frac{p^k+1}{2}}+b(-\pi^{\frac{p^k+1}{2}})^{i})\right)_{i=0}^{p^m-2}\mid a,b \in \mathbb{F}_{p^m}\right\},
\end{equation}
whose dual  has two zeros $\pi^{-\frac{p^k+1}{2}}$ and $-\pi^{\frac{p^k+1}{2}}$, where $k$ satisfies $\pi^{\frac{p^k+1}{2}p^i}\neq -\pi^{\frac{p^k+1}{2}}$ for all $i\in \mathbb{Z}_m$. Let $h_1(x)$ and $h_2(x)$ be the minimal polynomials of $\pi^{-\frac{p^k+1}{2}}$ and $-\pi^{-\frac{p^k+1}{2}}$ over $\mathbb{F}_p$, respectively. Then we have the following lemma.
\begin{lemma}\label{lem:degree}
With the  notations given above. The degree of $h_1(x)$ and $h_2(x)$ are both $m$.
\end{lemma}
\pf In order to calculate the degree of $h_1(x)$, we need to compute the size of $\Gamma_{\frac{p^k+1}{2}}$. Let the  size of $\Gamma_{\frac{p^k+1}{2}}$ be $l$. Note that $\frac{p^k+1}{2}p^m \equiv \frac{p^k+1}{2} \ ({\rm mod}\ p^m-1)$, then we get that $l \mid m$.
On the other hand, we have
\[ \frac{p^k+1}{2}p^l \equiv \frac{p^k+1}{2} \ ({\rm mod}\ p^m-1),\]
which implies that
\begin{equation}\label{eq:pm-1-pk+1}
 2(p^m-1) \mid (p^k+1)(p^l-1).
\end{equation}

Case I, when $v_2(k)\geq v_2(m)$: By Lemma~\ref{lem:gcdkm},  we have $\gcd(p^k+1,p^m-1)=2$ . From (\ref{eq:pm-1-pk+1}), we get $m\mid l$. Since $l\mid m$, hence, $m=l$.

Case II, when $ v_2(k)< v_2(m)$: In this case, we obtain that $\gcd(p^k+1,p^m-1)=p^d+1$ by Lemma~\ref{lem:gcdkm}. From (\ref{eq:pm-1-pk+1}), we have $2\frac{p^m-1}{p^d+1} \mid (p^l-1).$ Since $ v_2(k)< v_2(m)$, then $p^{2d}-1 \mid p^m-1$, which implies that $p^{d}-1 \mid 2\frac{p^m-1}{p^d+1}$. This shows that  $p^d-1\mid p^l-1$, i.e., $d\mid l$. Let $l=hd$, $m=sd$, where $h\mid s$ since $l \mid m$. From (\ref{eq:pm-1-pk+1}), we have \[2(p^{sd}-1) \mid (p^d+1)(p^{hd}-1), \]
which implies that  $h=s$. Then $l=m$.

Similarly, we also get that the degree of $h_2(x)$ is $m$.\EOP

From (\ref{eq:wtcab}), the  weight of $\textbf{c}(a,b)\in \mathcal{C}_{\frac{p^k+1}{2}}$ can be expressed as
\begin{equation}\label{eq:wt}
wt(\textbf{c}(a,b))=p^m-p^{m-1}-\frac{1}{2p}\sum_{u\in \mathbb{F}_p^*}\sum_{x\in \mathbb{F}_{p^m}}\left(\zeta_p^{u {\rm Tr}_1^m((a+b)x^{p^k+1})}+\zeta_p^{u {\rm Tr}_1^m((a-b)\pi^{\frac{p^k+1}{2}} x^{p^k+1})}\right).
\end{equation}

\subsection{The weight distribution of  $\mathcal{C}_{\frac{p^k+1}{2}}$ for $v_2(m)> v_2(k)$}
In this subsection, we always assume that $v_2(m)> v_2(k)$, i.e.,  $s=\frac{m}{d}$ is even, where $d=\gcd(m,k)$. Following the above notations, we let
\begin{equation}\label{eq:Sabeven}
T(a,b)=\sum_{u\in \mathbb{F}_p^*}\sum_{x\in \mathbb{F}_{p^m}}\left(\zeta_p^{u {\rm Tr}_1^m((a+b)x^{p^k+1})}+\zeta_p^{u {\rm Tr}_1^m((a-b)\pi^{\frac{p^k+1}{2}} x^{p^k+1})}\right).
\end{equation}
From (\ref{eq:wt}), the weight distribution of cyclic code $\mathcal{C}_{\frac{p^k+1}{2}}$ is completely
determined by the value distribution of $T(a,b)$.
To calculate the value distribution of $T(a,b)$, we need a series of lemmas. Before introducing them, we define
$$
R_\alpha(x)= \sum_{u\in \mathbb{F}_p^*}\sum_{x\in \mathbb{F}_{p^m}}\zeta_p^{u{\rm Tr}_1^m(\alpha x^{p^k+1} )},\  \alpha \in \mathbb{F}_{p^m}.
$$
\begin{lemma}\label{lem:Rx}
With the notations given above, we have
\begin{equation}\label{eq:Dxzuizhongxingshi}
R_\alpha(x)=(p-1)\sum_{x\in \mathbb{F}_{p^m}}\zeta_p^{{\rm Tr}_1^m(\alpha x^{p^k+1} )}.
\end{equation}
\end{lemma}
\pf Note that  $v_2(m)>v_2(k)$, then $m$ is even and $\frac{k}{d}$ is odd. Hence, $u_p=\pi^{\frac{p^m-1}{p-1}}=\pi^{p^{m-1}+\cdots+1}$ is a square element in $\mathbb{F}_{p^m}$, i.e., $u_p\in SQ$.
Since $\frac{k}{d}$ is odd and
$$u_p^{\frac{p^k+1}{2}}=u_p^{\frac{p^k-1}{2}+1}=u_pu_p^{\frac{p^d-1}{2}(p^{(\frac{k}{d}-1)d}+\cdots+1)}=u_p(u_p^{\frac{p-1}{2}})^{(p^{d-1}+\cdots+1)(p^{(\frac{k}{d}-1)d}+\cdots+1)},$$
we have that $u_p^{\frac{p^k+1}{2}}=-u_p$ for odd $d$, and $u_p^{\frac{p^k+1}{2}}=u_p$ for even $d$, i.e., $u_p^{\frac{p^k+1}{2}}=(-1)^d u_p$.
Using $u_p\in SQ$, we have
\begin{eqnarray}\label{eq:D1(x)}
\nonumber  R_\alpha(x) &=& \sum_{u\in \mathbb{F}_p^*}\sum_{x\in \mathbb{F}_{p^m}}\zeta_p^{u{\rm Tr}_1^m(\alpha x^{p^k+1} )} \\
\nonumber   &=& \sum_{u\in SQ_p}\sum_{x\in \mathbb{F}_{p^m}}\left(\zeta_p^{{\rm Tr}_1^m(\alpha (u^{\frac{1}{2}}x)^{p^k+1} )}+\zeta_p^{u_p{\rm Tr}_1^m(\alpha (u^{\frac{1}{2}}x)^{p^k+1} )}\right)\\
\nonumber   &=&  \frac{p-1}{2}\sum_{x\in \mathbb{F}_{p^m}}\left(\zeta_p^{{\rm Tr}_1^m(\alpha x^{p^k+1} )}+\zeta_p^{u_p{\rm Tr}_1^m(\alpha x^{p^k+1} )}\right)\\
\nonumber   &=&  \frac{p-1}{2}\sum_{x\in \mathbb{F}_{p^m}}\left(\zeta_p^{{\rm Tr}_1^m(\alpha x^{p^k+1} )}+\zeta_p^{{\rm Tr}_1^m((-1)^d\alpha (u_p^{\frac{1}{2}}x)^{p^k+1} )}\right)\\
\nonumber   &=&  \frac{p-1}{2}\sum_{x\in \mathbb{F}_{p^m}}\left(\zeta_p^{{\rm Tr}_1^m(\alpha x^{p^k+1} )}+\zeta_p^{{\rm Tr}_1^m((-1)^d\alpha x^{p^k+1} )}\right)\\
   &=& \left\{
         \begin{array}{ll}
           (p-1)\sum\limits_{x\in \mathbb{F}_{p^m}}\zeta_p^{{\rm Tr}_1^m(\alpha x^{p^k+1} )}, & \hbox{if $d$ is even;} \\
           \frac{p-1}{2}\sum\limits_{x\in \mathbb{F}_{p^m}}\left(\zeta_p^{{\rm Tr}_1^m(\alpha x^{p^k+1} )}+\zeta_p^{{\rm Tr}_1^m(-\alpha x^{p^k+1} )}\right), & \hbox{if $d$ is odd.}
         \end{array}
       \right.
\end{eqnarray}
Since  $ v_2(m)>v_2(k)$, by Lemma~\ref{lem:gcdkm}, we have $$\gcd(p^k+1,p^m-1)=p^{d}+1. $$
Note that $m=sd$ is even, then $2(p^d+1) \mid p^m-1$. Hence, we have $$(\pi^{\frac{p^m-1}{2(p^d+1)}})^{p^k+1}=(\pi^{\frac{p^m-1}{2}})^{\frac{p^k+1}{p^d+1}}=(-1)^{\frac{p^k+1}{p^d+1}}.$$
Since $\frac{k}{d}$ is odd, then  $\frac{p^k+1}{p^d+1}=p^{(\frac{k}{d}-1)d}-p^{(\frac{k}{d}-2)d}+p^{(\frac{k}{d}-3)d}-\cdots-p+1$ is odd, which implies that $(\pi^{\frac{p^m-1}{2(p^d+1)}})^{p^k+1}=-1$.
Therefore, we  have
\[\sum_{x\in \mathbb{F}_{p^m}}\zeta_p^{{\rm Tr}_1^m(\alpha x^{p^k+1} )} =\sum_{x\in \mathbb{F}_{p^m}}\zeta_p^{{\rm Tr}_1^m(\alpha (\pi^{\frac{p^m-1}{2(p^d+1)}}x)^{p^k+1} )}=\sum_{x\in \mathbb{F}_{p^m}}\zeta_p^{{\rm Tr}_1^m(-\alpha x^{p^k+1} )}.\]
By (\ref{eq:D1(x)}), we obtain
$$
R_\alpha(x)=(p-1)\sum_{x\in \mathbb{F}_{p^m}}\zeta_p^{{\rm Tr}_1^m(\alpha x^{p^k+1} )}.
$$\EOP

Applying Lemmas~\ref{lemTfenbu} and \ref{lem:Rx}, we have the following result.
\begin{lemma}\label{lemDxfenbu2}
With the notations given above.
\begin{itemize}
  \item If $v_2(k)+1=v_2(m)$, then
  \[R_\alpha(x)=\left\{
    \begin{array}{ll}
     (p-1)p^m, & \hbox{$1$ time;} \\
      -(p-1)p^{\frac{m}{2}}, & \hbox{$\frac{p^d(p^m-1)}{p^d+1}$ times;}\\
      (p-1) p^{\frac{m}{2}+d}, & \hbox{$\frac{p^m-1}{p^d+1}$ times.}
    \end{array}
  \right.\]
  \item If $v_2(k)+1<v_2(m)$, then
        \[R_\alpha(x)=\left\{
    \begin{array}{ll}
     (p-1)p^m, & \hbox{$1$ time;} \\
      (p-1)p^{\frac{m}{2}}, & \hbox{$\frac{p^d(p^m-1)}{p^d+1}$ times;}\\
      -(p-1) p^{\frac{m}{2}+d}, & \hbox{$\frac{p^m-1}{p^d+1}$ times.}
    \end{array}
  \right.\]
\end{itemize}
\end{lemma}
\begin{lemma}\label{lem:Squzhi}
With the notations given above. Suppose $(a,b)$ runs through $\mathbb{F}_{p^m}^2$,
\begin{itemize}
  \item if $v_2(k)+1=v_2(m)$, then $T(a,b)$ takes  on only  the values from  set of $$\{2(p-1)p^m, (p-1)(p^m+ p^{\frac{m}{2}+d}), (p-1)( p^{m}-p^{\frac{m}{2}}), (p-1)( p^{\frac{m}{2}+d}-p^{\frac{m}{2}}),2(p-1) p^{\frac{m}{2}+d},-2(p-1) p^{\frac{m}{2}}\}.$$
  \item if $v_2(k)+1<v_2(m)$,
  then $T(a,b)$ takes  on only  the values from  set of $$\{2(p-1)p^m, (p-1)(p^m-p^{\frac{m}{2}+d}),(p-1)(p^m+p^{\frac{m}{2}}),(p-1)(p^{\frac{m}{2}}- p^{\frac{m}{2}+d}),-2(p-1) p^{\frac{m}{2}+d},2(p-1) p^{\frac{m}{2}}   \}.$$
\end{itemize}
\end{lemma}
\pf By (\ref{eq:Sabeven}) and (\ref{eq:Dxzuizhongxingshi}), we have
$$T(a,b)=R_{a+b}(x)+R_{(a-b)\pi^{\frac{p^k+1}{2}}}(x).$$

We first discuss the value of $T(a,b)$ in the case of  $v_2(k)+1=v_2(m)$.

Case I, when $a=b=0$: It is easy to check that $T(a,b)=2(p-1)p^m$.

Case II, when $a=b\neq 0$ or $a=-b\neq 0$: We first discuss the case of $a=b\neq 0$, since the other case can be discussed by a similar way. By Lemma~\ref{lemDxfenbu2}, we have $R_{a+b}(x)\in \{(p-1) p^{\frac{m}{2}+d},-(p-1)p^{\frac{m}{2}}\}$, and $R_{(a-b)\pi^{\frac{p^k+1}{2}}}(x)=(p-1)p^m $. Hence, $T(a,b)\in \{(p-1)(p^m+p^{\frac{m}{2}+d}), (p-1)(p^m-p^{\frac{m}{2}}) \}$. Similarly, we get $T(a,b)\in \{(p-1)(p^m+p^{\frac{m}{2}+d}), (p-1)(p^m-p^{\frac{m}{2}}) \}$ in the case of $a=-b\neq 0$.

Case III, when $a\neq b $, $a \neq -b $: By Lemma~~\ref{lemDxfenbu2}, we have $R_{a+b}(x)\in \{(p-1) p^{\frac{m}{2}+d},-(p-1)p^{\frac{m}{2}}\}$ and $R_{(a-b)\pi^{\frac{p^k+1}{2}}}(x)\in \{(p-1) p^{\frac{m}{2}+d},-(p-1)p^{\frac{m}{2}}\}$. Therefore, $T(a,b)\in \{2(p-1) p^{\frac{m}{2}+d},-2(p-1)p^{\frac{m}{2}}, (p-1)(p^{\frac{m}{2}+d}-p^{\frac{m}{2}})\}$.

The case of  $v_2(k)+1<v_2(m)$ can be discussed by a similar way as  the case of  $v_2(k)+1=v_2(m)$. This completes the proof.\EOP\\

With above preparation we can determine the value distribution of the exponential sum $T(a,b)$ defined by (\ref{eq:Sabeven}).
\begin{theorem}\label{th:Txfenbu}
With the notations given above. Suppose $(a, b)$ runs
through $\mathbb{F}_{p^m}^2$,
\begin{itemize}
  \item if $v_2(k)+1=v_2(m)$, then the value distribution of $T(a,b)$ is given as follows.
  $$\left\{
    \begin{array}{ll}
      2(p-1)p^m, & \hbox{$1$ time;} \\
      (p-1)(p^m+ p^{\frac{m}{2}+d}), & \hbox{$2\frac{p^m-1}{p^d+1}$ times;} \\
      (p-1)( p^{m}-p^{\frac{m}{2}}), & \hbox{$2\frac{p^d(p^m-1)}{p^d+1}$ times;} \\
      (p-1)( p^{\frac{m}{2}+d}-p^{\frac{m}{2}}), & \hbox{$2p^d(\frac{p^m-1}{p^d+1}) ^2$times;} \\
      2(p-1) p^{\frac{m}{2}+d}, & \hbox{$(\frac{p^m-1}{p^d+1}) ^2 $times;} \\
      -2(p-1) p^{\frac{m}{2}}, & \hbox{$p^{2d}(\frac{p^m-1}{p^d+1}) ^2$times.}
    \end{array}
  \right.$$
  \item if $v_2(k)+1<v_2(m)$, then the value distribution of $T(a,b)$ is given as follows.
  $$\left\{
    \begin{array}{ll}
      2(p-1)p^m, & \hbox{$1$ time;} \\
      (p-1)(p^m- p^{\frac{m}{2}+d}), & \hbox{$2\frac{p^m-1}{p^d+1}$ times;} \\
      (p-1)( p^{m}+p^{\frac{m}{2}}), & \hbox{$2\frac{p^d(p^m-1)}{p^d+1}$ times;} \\
      (p-1)( p^{\frac{m}{2}}-p^{\frac{m}{2}+d}), & \hbox{$2p^d(\frac{p^m-1}{p^d+1}) ^2$times;} \\
      -2(p-1) p^{\frac{m}{2}+d}, & \hbox{$(\frac{p^m-1}{p^d+1}) ^2 $times;} \\
      2(p-1) p^{\frac{m}{2}}, & \hbox{$p^{2d}(\frac{p^m-1}{p^d+1}) ^2$times.}
    \end{array}
  \right.$$
\end{itemize}

\end{theorem}
\pf We only discuss the case of $v_2(k)+1=v_2(m)$, since the other case can be discussed by a similar way.
To determine the value distribution of $T(a,b)$, we define
$$N_1=\#\{a,b\in \mathbb{F}_{p^m}\mid T(a,b)=(p-1)(p^m+ p^{\frac{m}{2}+d})\},N_2=\#\{a,b\in \mathbb{F}_{p^m}\mid T(a,b)= (p-1)( p^{m}-p^{\frac{m}{2}})\}, $$
$$N_3=\#\{a,b\in \mathbb{F}_{p^m}\mid T(a,b)=(p-1)( p^{\frac{m}{2}+d}-p^{\frac{m}{2}})\},N_4=\#\{a,b\in \mathbb{F}_{p^m}\mid T(a,b)=2(p-1) p^{\frac{m}{2}+d}\},$$ $$N_5=\#\{a,b\in \mathbb{F}_{p^m}\mid T(a,b)=-2(p-1) p^{\frac{m}{2}}\}.$$
It is easy to check that the value   $2(p-1)p^m$ happens only once. Note that the value $(p-1)(p^m+ p^{\frac{m}{2}+d})$ occurs only if $R_{a+b}(x)= (p-1)p^m$, $R_{(a-b)\pi^{\frac{p^k+1}{2}}}(x)=(p-1) p^{\frac{m}{2}+d}$ or $R_{a+b}(x)= (p-1) p^{\frac{m}{2}+d}$, $R_{(a-b)\pi^{\frac{p^k+1}{2}}}(x)=(p-1)p^m$. By Lemma~\ref{lemDxfenbu2}, if $R_{a+b}(x)= (p-1)p^m$, $R_{(a+b)\pi^{\frac{p^k+1}{2}}}(x)=(p-1) p^{\frac{m}{2}+d}$, then we have that the number of $(a,b)\in \mathbb{F}_{p^m}^2$ is $\frac{p^m-1}{p^d+1}$. By Lemma~\ref{lemDxfenbu2}, if $R_{a+b}(x)= (p-1) p^{\frac{m}{2}+d}$, $R_{(a-b)\pi^{\frac{p^k+1}{2}}}(x)=(p-1)p^m$, we get that the number of $(a,b)\in \mathbb{F}_{p^m}^2$ is $\frac{p^m-1}{p^d+1}$.  Therefore, we have $N_1=2\frac{p^m-1}{p^d+1}$.  Similarly, we obtain that $N_2=2\frac{p^d(p^m-1)}{p^d+1}$. On the other hand, we have
\begin{eqnarray*}
  N_3 &=& \#\{a,b\in \mathbb{F}_{p^m}\mid T(a,b)=R_{a+b}(x)+R_{(a-b)\pi^{\frac{p^k+1}{2}}}(x)=(p-1)( p^{\frac{m}{2}+d}-p^{\frac{m}{2}})\} \\
   &=& \#\{u,v\in \mathbb{F}_{p^m}\mid R_{u}(x)+R_{v}(x)=(p-1)( p^{\frac{m}{2}+d}-p^{\frac{m}{2}})\} \\
   &=& \#\{u,v\in \mathbb{F}_{p^m}\mid R_{u}(x)=(p-1) p^{\frac{m}{2}+d},R_{v}(x)=-(p-1)p^{\frac{m}{2}}\} \\
   &&+ \#\{u,v\in \mathbb{F}_{p^m}\mid R_{v}(x)=(p-1) p^{\frac{m}{2}+d},R_{u}(x)=-(p-1)p^{\frac{m}{2}}\}.
\end{eqnarray*}
By Lemma~\ref{lemDxfenbu2}, we obtain $N_3=2p^d(\frac{p^m-1}{p^d+1}) ^2.$
Similarly, we get
\[N_4=(\frac{p^m-1}{p^d+1}) ^2,\ \ \ N_5=p^{2d}(\frac{p^m-1}{p^d+1}) ^2.\]
This finishes the proof.
\EOP
     \begin{table}[!h]
\tabcolsep 0pt
\caption{for the case of  $v_2(k)+1=v_2(m)$}
\vspace*{-12pt}
\begin{center}
\def\temptablewidth{0.5\textwidth}
{\rule{\temptablewidth}{1pt}}
\begin{tabular*}{\temptablewidth}{@{\extracolsep{\fill}}cc}
Weight & Frequency  \\   \hline
  $0$ & $1$ \\
  $\frac{p-1}{2}(p^{m-1}- p^{\frac{m}{2}+d-1})$ & $2\frac{p^m-1}{p^d+1}$ \\
  $\frac{p-1}{2}( p^{m-1}+p^{\frac{m}{2}-1})$ & $2\frac{p^d(p^m-1)}{p^d+1}$ \\
  $\frac{p-1}{2}(2p^{m-1}- p^{\frac{m}{2}+d-1}+p^{\frac{m}{2}-1})$ & $2p^d(\frac{p^m-1}{p^d+1}) ^2$ \\
    $(p-1)(p^{m-1}- p^{\frac{m}{2}+d-1})$ & $(\frac{p^m-1}{p^d+1}) ^2 $ \\
  $(p-1)(p^{m-1}+ p^{\frac{m}{2}-1})$ & $p^{2d}(\frac{p^m-1}{p^d+1}) ^2$\\
      \end{tabular*}
       {\rule{\temptablewidth}{1pt}}
       \end{center}
       \end{table}
\begin{theorem}\label{th:ceven}
With the notations given above.
\begin{itemize}
  \item If $v_2(k)+1=v_2(m)$, then  $\mathcal{C}_{\frac{p^k+1}{2}}$ is a cyclic code over $\mathbb{F}_p$ with parameters $[p^m-1,2m,\frac{p-1}{2}(p^{m-1}- p^{\frac{m}{2}+d-1})]$ and the  weight distribution of  $\mathcal{C}_{\frac{p^k+1}{2}}$ is given in Table $1$.

  \item If $v_2(k)+1<v_2(m)$, then $\mathcal{C}_{\frac{p^k+1}{2}}$ is a cyclic code over $\mathbb{F}_p$ with parameters $[p^m-1,2m,\frac{p-1}{2}( p^{m-1}-p^{\frac{m}{2}-1})]$ and the  weight distribution of  $\mathcal{C}_{\frac{p^k+1}{2}}$ is given in Table $2$.
      \begin{table}[!h]
\tabcolsep 0pt
\caption{for the case of  $v_2(k)+1<v_2(m)$}
\vspace*{-12pt}
\begin{center}
\def\temptablewidth{0.5\textwidth}
{\rule{\temptablewidth}{1pt}}
\begin{tabular*}{\temptablewidth}{@{\extracolsep{\fill}}cc}
  Weight & Frequency\\ \hline
  $0$ & $1$ \\
  $\frac{p-1}{2}(p^{m-1}+ p^{\frac{m}{2}+d-1})$ & $2\frac{p^m-1}{p^d+1}$ \\
  $\frac{p-1}{2}( p^{m-1}-p^{\frac{m}{2}-1})$ & $2\frac{p^d(p^m-1)}{p^d+1}$ \\
  $\frac{p-1}{2}(2p^{m-1}+ p^{\frac{m}{2}+d-1}-p^{\frac{m}{2}-1})$ & $2p^d(\frac{p^m-1}{p^d+1}) ^2$ \\
    $(p-1)(p^{m-1}+ p^{\frac{m}{2}+d-1})$ & $(\frac{p^m-1}{p^d+1}) ^2 $ \\
  $(p-1)(p^{m-1}- p^{\frac{m}{2}-1})$ & $p^{2d}(\frac{p^m-1}{p^d+1}) ^2$\\
       \end{tabular*}
       {\rule{\temptablewidth}{1pt}}
       \end{center}
       \end{table}
\end{itemize}
\end{theorem}

\pf Combining Theorem~\ref{th:Txfenbu}, Lemma~\ref{lem:degree} and  (\ref{eq:wt}), we finish the proof.\EOP
\subsection{The weight distribution of  $\mathcal{C}_{\frac{p^k+1}{2}}$ for $v_2(m)\leq v_2(k)$}
In this subsection, we always assume that $v_2(m)\leq v_2(k)$, i.e.,  $s=\frac{m}{d}$ is odd, where $d=\gcd(m,k)$.
\begin{lemma}\label{lem:C1}
With the notations given above,  the codes $\mathcal{C}_1$ and $\mathcal{C}_{\frac{p^k+1}{2}}$ have the same weight distribution.
\end{lemma}
\pf By (\ref{eq:wtcab}), we have that the weight distribution of $\mathcal{C}_1$ and $\mathcal{C}_{\frac{p^k+1}{2}}$ are respectively determined by the value distribution of \[\Delta_1= \sum_{u\in \mathbb{F}_p^*}\sum_{x\in \mathbb{F}_{p^m}}\left(\zeta_p^{u {\rm Tr}_1^m((a+b)x^{2})}+\zeta_p^{u {\rm Tr}_1^m((a-b)\pi x^{2})}\right)\]
and \[\Delta_{\frac{p^k+1}{2}}=\sum_{u\in \mathbb{F}_p^*}\sum_{x\in \mathbb{F}_{p^m}}\left(\zeta_p^{u {\rm Tr}_1^m((a+b)x^{p^k+1})}+\zeta_p^{u {\rm Tr}_1^m((a-b)\pi^{\frac{p^k+1}{2}} x^{p^k+1})}\right).\]
Since $v_2(k)\geq v_2(m)$, by Lemma~\ref{lem:gcdkm}, we have $\gcd(p^m-1,p^k+1)=2$, which implies that $\{x^{p^k+1}\mid x\in \mathbb{F}_{p^m}\}=\{x^2\mid x\in \mathbb{F}_{p^m}\}.$ Hence
\begin{eqnarray*}
  \Delta_{\frac{p^k+1}{2}}&= & \sum_{u\in \mathbb{F}_p^*}\sum_{x\in \mathbb{F}_{p^m}}\left(\zeta_p^{u {\rm Tr}_1^m((a+b)x^{p^k+1})}+\zeta_p^{u {\rm Tr}_1^m((a-b)\pi^{\frac{p^k+1}{2}} x^{p^k+1})}\right) \\
   &=& \sum_{u\in \mathbb{F}_p^*}\sum_{x\in \mathbb{F}_{p^m}}\left(\zeta_p^{u {\rm Tr}_1^m((a+b)x^{2})}+\zeta_p^{u {\rm Tr}_1^m((a-b)\pi^{\frac{p^k+1}{2}} x^{2})}\right)\\
   &=&\left\{
        \begin{array}{ll}
          \sum\limits_{u\in \mathbb{F}_p^*}\sum\limits_{x\in \mathbb{F}_{p^m}}\left(\zeta_p^{u {\rm Tr}_1^m((a+b)x^{2})}+\zeta_p^{u {\rm Tr}_1^m((a-b) x^{2})}\right), & \hbox{ if $p^k\equiv 3 \ ({\rm mod }\ 4) $,} \\ \\
          \sum\limits_{u\in \mathbb{F}_p^*}\sum\limits_{x\in \mathbb{F}_{p^m}}\left(\zeta_p^{u {\rm Tr}_1^m((a+b)x^{2})}+\zeta_p^{u {\rm Tr}_1^m((a-b)\pi x^{2})}\right), & \hbox{ if $p^k\equiv 1 \ ({\rm mod }\ 4) $.}
        \end{array}
      \right.
\end{eqnarray*}
If $p^k\equiv 1 \ ({\rm mod }\ 4) $, then $\Delta_1=\Delta_{\frac{p^k+1}{2}}$. On the other hand, if $p^k\equiv 3 \ ({\rm mod }\ 4) $, then $k$ is odd. Since $s$ is odd, then $m$ is odd, which implies that $u_p$ is a nonsquare element in $\mathbb{F}_{p^m}$. In the following, we only need to prove that
\[ \sum_{u\in \mathbb{F}_p^*}\sum_{x\in \mathbb{F}_{p^m}}\zeta_p^{u {\rm Tr}_1^m((a-b)x^{2})}=\sum_{u\in \mathbb{F}_p^*}\sum_{x\in \mathbb{F}_{p^m}}\zeta_p^{u {\rm Tr}_1^m((a-b)\pi x^{2})}.\]
On the other hand, we have that
\begin{eqnarray}\label{eq:3.8}
\nonumber   & & \sum_{u\in \mathbb{F}_p^*}\sum_{x\in \mathbb{F}_{p^m}}\zeta_p^{u {\rm Tr}_1^m((a-b)x^{2})} \\
\nonumber   &=& \sum_{u\in SQ_p}\sum_{x\in \mathbb{F}_{p^m}}\left(\zeta_p^{ {\rm Tr}_1^m((a-b)(u^{\frac{1}{2}}x)^{2})}+\zeta_p^{ {\rm Tr}_1^m(u_p(a-b)(u^{\frac{1}{2}}x)^{2})}\right)\\
\nonumber   &=&\frac{p-1}{2}\sum_{x\in \mathbb{F}_{p^m}}\left(\zeta_p^{ {\rm Tr}_1^m((a-b)x^{2})}+\zeta_p^{{\rm Tr}_1^m(u_p(a-b)x^{2})}\right)\\
\nonumber  &=&\frac{p-1}{2}\sum_{x\in \mathbb{F}_{p^m}}\left(\zeta_p^{ {\rm Tr}_1^m((a-b)x^{2})}+\zeta_p^{ {\rm Tr}_1^m(\pi^{\frac{p^m-1}{p-1}}(a-b)x^{2})}\right)\\
\nonumber   &=&\frac{p-1}{2}\sum_{x\in \mathbb{F}_{p^m}}\left(\zeta_p^{ {\rm Tr}_1^m((a-b)x^{2})}+\zeta_p^{ {\rm Tr}_1^m( (a-b)\pi(\pi^{\frac{p^{m-1}+p^{m-2}+\cdots+p}{2}}x)^{2})}\right)\\
   &=&\frac{p-1}{2}\sum_{x\in \mathbb{F}_{p^m}}\left(\zeta_p^{ {\rm Tr}_1^m((a-b)x^{2})}+\zeta_p^{{\rm Tr}_1^m((a-b)\pi x^{2})}\right),
\end{eqnarray}
and
\begin{eqnarray}\label{eq:3.9}
\nonumber   & & \sum_{u\in \mathbb{F}_p^*}\sum_{x\in \mathbb{F}_{p^m}}\zeta_p^{u {\rm Tr}_1^m((a-b)\pi x^{2})} \\
\nonumber   &=& \sum_{u\in SQ_p}\sum_{x\in \mathbb{F}_{p^m}}\left(\zeta_p^{ {\rm Tr}_1^m((a-b)\pi(u^{\frac{1}{2}}x)^{2})}+\zeta_p^{ {\rm Tr}_1^m(u_p(a-b)\pi(u^{\frac{1}{2}}x)^{2})}\right)\\
\nonumber   &=&\frac{p-1}{2}\sum_{x\in \mathbb{F}_{p^m}}\left(\zeta_p^{ {\rm Tr}_1^m((a-b)\pi x^{2})}+\zeta_p^{ {\rm Tr}_1^m(u_p(a-b)\pi x^{2})}\right)\\
\nonumber  &=&\frac{p-1}{2}\sum_{x\in \mathbb{F}_{p^m}}\left(\zeta_p^{ {\rm Tr}_1^m((a-b)\pi x^{2})}+\zeta_p^{ {\rm Tr}_1^m(\pi^{\frac{p^m-1}{p-1}+1}(a-b)x^{2})}\right)\\
\nonumber   &=&\frac{p-1}{2}\sum_{x\in \mathbb{F}_{p^m}}\left(\zeta_p^{ {\rm Tr}_1^m((a-b)\pi x^{2})}+\zeta_p^{ {\rm Tr}_1^m( (a-b)(\pi^{\frac{p^{m-1}+p^{m-2}+\cdots+p+2}{2}}x)^{2})}\right)\\
   &=&\frac{p-1}{2}\sum_{x\in \mathbb{F}_{p^m}}\left(\zeta_p^{ {\rm Tr}_1^m((a-b)\pi x^{2})}+\zeta_p^{{\rm Tr}_1^m((a-b) x^{2})}\right).
\end{eqnarray}
By comparing (\ref{eq:3.8}) and (\ref{eq:3.9}), we finish the proof.\EOP

By Lemma~\ref{lem:C1}, the weight distribution of the code $\mathcal{C}_{\frac{p^k+1}{2}}$ is the same as the code $\mathcal{C}_1$. As we know, the weight distribution of the code $\mathcal{C}_1$  has been studied in \cite{Ma}( see  Theorems~5,6 ).

\begin{lemma}\label{lem:C1weight}{\rm \cite{Ma}}
With the notations given above.
\begin{itemize}
  \item If $v_2(m)=0$, then $\mathcal{C}_1$ is a cyclic code over $\mathbb{F}_p$ with parameters $[p^m-1,2m,\frac{p-1}{2}p^{m-1}]$ and the  weight distribution of  $\mathcal{C}_{\frac{p^k+1}{2}}$ is given in Table $3$.
  \item If $1\leq v_2(m)\leq v_2(k)$, then $\mathcal{C}_1$ is a cyclic code over $\mathbb{F}_p$ with parameters $[p^m-1,2m,\frac{p-1}{2}(p^{m-1}-p^{\frac{m}{2}-1})]$ and the  weight distribution of  $\mathcal{C}_{\frac{p^k+1}{2}}$ is given in Table $4$.
\end{itemize}
\end{lemma}
\begin{table}[!h]
\tabcolsep 0pt
\caption{for the case of $v_2(m)=0$}
\vspace*{-12pt}
\begin{center}
\def\temptablewidth{0.5\textwidth}
{\rule{\temptablewidth}{1pt}}
\begin{tabular*}{\temptablewidth}{@{\extracolsep{\fill}}cc}
Weight & Frequency  \\   \hline
    $0$ & $1$ \\
  $\frac{p-1}{2}p^{m-1}$ & $2(p^m-1)$ \\
  $(p-1)p^{m-1}$ & $p^{2m}-2p^m+1$ \\
       \end{tabular*}
       {\rule{\temptablewidth}{1pt}}
       \end{center}
       \end{table}
\begin{table}[!h]
\tabcolsep 0pt
\caption{for the case of $1\leq v_2(m)\leq v_2(k)$}
\vspace*{-12pt}
\begin{center}
\def\temptablewidth{0.5\textwidth}
{\rule{\temptablewidth}{1pt}}
\begin{tabular*}{\temptablewidth}{@{\extracolsep{\fill}}cc}
Weight & Frequency  \\   \hline
  $0$ & $1$ \\
  $(p-1)p^{m-1}$ & $\frac{(p^m-1)^2}{2}$ \\
  $(p-1)(p^{m-1}+p^{\frac{m}{2}-1})$ & $\frac{(p^m-1)^2}{4}$ \\
  $(p-1)(p^{m-1}-p^{\frac{m}{2}-1})$ & $\frac{(p^m-1)^2}{4}$ \\
  $\frac{p-1}{2}(p^{m-1}+p^{\frac{m}{2}-1})$ & $p^m-1$ \\
  $\frac{p-1}{2}(p^{m-1}-p^{\frac{m}{2}-1})$ & $p^m-1$ \\
       \end{tabular*}
       {\rule{\temptablewidth}{1pt}}
       \end{center}
       \end{table}
Applying Lemmas~\ref{lem:C1} and \ref{lem:C1weight}, we have the following theorem.
\begin{theorem}\label{th:codd}
With the notations given above.
\begin{itemize}
  \item If $v_2(m)=0$, then $\mathcal{C}_{\frac{p^k+1}{2}}$ is a cyclic code over $\mathbb{F}_p$ with parameters $[p^m-1,2m,\frac{p-1}{2}p^{m-1}]$ and the  weight distribution of  $\mathcal{C}_{\frac{p^k+1}{2}}$ is given in Table $3$.
  \item If $1\leq v_2(m)\leq v_2(k)$, then $\mathcal{C}_{\frac{p^k+1}{2}}$ is a cyclic code over $\mathbb{F}_p$ with parameters $[p^m-1,2m,\frac{p-1}{2}(p^{m-1}-p^{\frac{m}{2}-1})]$ and the  weight distribution of  $\mathcal{C}_{\frac{p^k+1}{2}}$ is given in Table $4$.
\end{itemize}
\end{theorem}

\subsection{The weight distribution of the code  $\mathcal{C}_t$ }
Combining Lemma~\ref{lem:zongfenlei}, Theorems~\ref{th:ceven} and ~\ref{th:codd}, we have the following main result in this paper.
\begin{theorem}\label{th:oddmain}
With the notations given above. Let  $t \in \mathbb{Z}_{p^m-1}$  be a  positive integer such that  $t\equiv \frac{p^k+1}{2} p^\tau \  ({\rm mod}\ \frac{p^m-1}{2})$ for some $\tau\in \mathbb{Z}_m$, where $k$ satisfies $\pi^{\frac{p^k+1}{2}p^i}\neq -\pi^{\frac{p^k+1}{2}}$ for all $i\in \mathbb{Z}_m$.
\begin{itemize}
    \item If  $v_2(k)+1=v_2(m)$, then $\mathcal{C}_t$ is a cyclic code over $\mathbb{F}_p$ with parameters $[p^m-1,2m,\frac{p-1}{2}(p^{m-1}- p^{\frac{m}{2}+d-1})]$ and the  weight distribution of  $\mathcal{C}_t$ is given by Table $1$.
  \item If  $v_2(k)+1<v_2(m)$, then $\mathcal{C}_t$ is a cyclic code over $\mathbb{F}_p$ with parameters $[p^m-1,2m,\frac{p-1}{2}(p^{m-1}-p^{\frac{m}{2}-1})]$ and the  weight distribution of  $\mathcal{C}_t$ is given by Table $2$.
  \item If $v_2(m)=0$, then $\mathcal{C}_t$ is a cyclic code over $\mathbb{F}_p$ with parameters $[p^m-1,2m,\frac{p-1}{2}p^{m-1}]$ and the  weight distribution of  $\mathcal{C}_t$ is given by Table $3$.
  \item If $1\leq v_2(m)\leq v_2(k)$, then $\mathcal{C}_t$ is a cyclic code over $\mathbb{F}_p$ with parameters $[p^m-1,2m,\frac{p-1}{2}(p^{m-1}-p^{\frac{m}{2}-1})]$ and the  weight distribution of  $\mathcal{C}_t$ is given by Table $4$.
\end{itemize}
\end{theorem}

In the following, we give two examples to verify our main results in Theorem~\ref{th:oddmain}.
\begin{example}
Let $p=3$, $m=6$, $k=1$. If $t\equiv 2 p^\tau \  ({\rm mod}\ 364)$ for some $\tau\in \mathbb{Z}_6$, then the code $\mathcal{C}_t$ is a $[728,12,216]$ cyclic
code over $\mathbb{F}_3$ with weight enumerator
\[1+364X^{216}+1092X^{252}+33124X^{432}+198744X^{468}+298116X^{504},\]
which confirms the weight distribution in Table $1$.
\end{example}
\begin{example}
Let $p=3$, $m=4$, $k=1$. If $t\equiv 2 p^\tau \  ({\rm mod}\ 40)$ for some $\tau\in \mathbb{Z}_4$, then the code $\mathcal{C}_t$ is a $[80,8,24]$ cyclic
code over $\mathbb{F}_3$ with weight enumerator
\[1+120X^{24}+40X^{36}+3600X^{48}+2400X^{60}+400X^{72},\]
which confirms the weight distribution in Table $2$.
\end{example}
\begin{example}
Let $p=5$, $m=3$, $k=1$. If $t\equiv 3 p^\tau \  ({\rm mod}\ 62)$ for some $\tau\in \mathbb{Z}_3$, then the code $\mathcal{C}_t$ is a $[124,6,50]$ cyclic
code over $\mathbb{F}_5$ with weight enumerator
\[1+248X^{50}+15376X^{100},\]
which confirms the weight distribution in Table $3$.
\end{example}
\begin{example}
Let $p=3$, $m=6$, $k=2$. If $t\equiv 5 p^\tau \  ({\rm mod}\ 364)$ for some $\tau\in \mathbb{Z}_6$, then the code $\mathcal{C}_t$ is a $[728,12,234]$ cyclic
code over $\mathbb{F}_3$ with weight enumerator
\[1+728X^{234}+728X^{252}+132496X^{468}+264992X^{486}+132496X^{504},\]
which confirms the weight distribution in Table $4$.
\end{example}


\begin{thebibliography}{99}
\bibitem{Delsarte} P. Delsarte, ``On subfield subcodes of modified Reed-Solomon codes,'' \emph{IEEE Trans. Inform. Theory}, \textbf{21}(5), 575-576(1975).


\bibitem{Ding2011} C. Ding,  Y. Liu, C. Ma, L. Zeng, ``The weight distributions of the duals of cyclic codes with two zeros,''
\emph{IEEE Trans. Inform. Theory}, \textbf{57}(12), 8000-8006(2011).

\bibitem{Ding2013} C. Ding,  J. Yang, ``Hamming weight in irrecducible codes,'' \emph{Discret. Math}. \textbf{313}(4), 434-446(2013).

\bibitem{Draper} S. Draper, X. Hou, ``explicit evaluation of certain exponential sums
of quadratic functions over $\mathbb{F}_{p^m}$, $p$ odd,'' http://arxiv.org/pdf/0708.3619v1.pdf.

\bibitem{FengandLuo}K. Feng, J. Luo, ``Weight distribution of some reducible cyclic codes,''\emph{Finite Fields Appl.}  \textbf{14}(2), 390-409(2008).


\bibitem{Feng} T. Feng, ``On cyclic codes of length $2^{2^r}-1$ with two zeros whose dual codes have three weights,'' \emph{Des. Codes Cryptogr.} \textbf{62}(3), 253-258(2012).

\bibitem{Helleseth-Kholosha2006} T. Helleseth, A. Kholosha, ``Monomial and quadratic bent functions over the finite fields of odd characteristic'', {\it
 IEEE Trans. Inform. Theory,}   {\bf 52}(5),  2018-2032(2006).

\bibitem{Klapper}A. Klapper, ``Cross-correlations of quadratic form sequences in odd characteristic,''\emph{Des. Codes Cryptogr.}  \textbf{3}(4), 289-305(1997).

\bibitem{Lichulei} C. Li, N. Li, T. Helleseth, C. Ding, ``On the weight distributions of several classes of cyclic codes from APN monomials,''Available: http://arxiv.org/pdf/1308.5885v1.pdf.


\bibitem{Lidl R} R. Lidl,  H. Niederreiter, ``Finite Fields,'' Ency clopedia of Mathematics,  \textbf{20}, Cambridge University Press,
Cambridge(1983).

\bibitem{Liu} Y. Liu, H. Yan, C. Liu, ``A class of six-weight cyclic codes and their weight distribution,'' Available: http://arxiv.org/pdf/1311.3391v2.pdf.

\bibitem{LuoandFeng1} J. Luo, K. Feng, ``Cyclic codes and sequences from generalized Coulter-Matthews function,'' \emph{IEEE Trans. Inform. Theory},  \textbf{54}(12), 5345-5353(2008).

\bibitem{LuoandFeng2}J. Luo, K. Feng, ``On the weight distributions of two classes of cyclic codes,'' \emph{IEEE Trans. Inform. Theory}, \textbf{54}(12), 5332-5344(2008).
\bibitem{Ma} C. Ma, L. zeng, Y. Liu, D. Feng, C. Ding, ``The weight enumerator of a class of cyclic codes,''
\emph{IEEE Trans. Inform. Theory}, \textbf{57}(1), 397-402(2011).

\bibitem{macwilliams} F. MacWilliams,  N. Sloane, ``The Theory of Error-Correcting Codes,'' North-Holland Publishing, Amsterdam(1997).
\bibitem{Rao} A. Rao, N. Pinnawala, ``A family of two-weight irreducible cyclic codes,'' \emph{IEEE Trans. Inform. Theory}, \textbf{56}(6), 2568-2570(2010).
\bibitem{Trachtenberg} H. Trachtenberg, ``On the crosscorrelation functions of maximal linear recurring sequences,'' Ph.D. dissertation, Univ. South. Calif.,
Los Angels(1970).












\bibitem{Vega12}G. Vega, ``The weight distribution of an extended class of reducible cyclic codes,'' \emph{IEEE Trans. Inform. Theory}, \textbf{58}(7), 4862-4869(2012).

\bibitem{Vega07}G. Vega, J. Wolfmann, ``New classes of $2$-weight cyclic codes,'' \emph{Des. Codes Cryptogr.} \textbf{42}(3), 327-334(2007).

\bibitem{Wang} B. Wang,  C. Tang, Y. Qi , Y. Yang , M. Xu, ``The weight distributions of cyclic codes and elliptic curves,''
\emph{IEEE Trans. Inform. Theory}, \textbf{58}(12), 7253-7259(2012).

\bibitem{XiongDCC}M. Xiong, ``The weight distributions of a class of cyclic codes II,'' \emph{Des. Codes Cryptogr.} (2012). doi: 10.1007/s10623-012-9785-0.

\bibitem{Xiong2013}M. Xiong, ``The weight distributions of a class of cyclic codes III,'' \emph{Finite Fields Appl.} \textbf{21}, 84-96(2013).

\bibitem{Xiong2012}M. Xiong, ``The weight distributions of a class of cyclic codes,'' \emph{Finite Fields Appl.} \textbf{18}(5), 933-945(2012).

\bibitem{Zeng2012} X. Zeng,  J. Shan,  L. Hu, ``A triple-error-correcting cyclic code from the Gold and Kasami-Welch APN
power functions,'' \emph{Finite Fields Appl.} \textbf{18}(1), 70-92(2012).

\bibitem{Zeng} X. Zeng,  L. Hu,  W. Jiang,  Q. Yue,  X. Cao, ``The weight distribution of a class of $p$-ary cyclic codes,'' \emph{Finite
Fields Appl.} \textbf{16}(1), 56-73(2010).

\bibitem{zheng} D. Zheng, X. Wang,
X. Zeng, L. Hu, ``The weight distribution of a family of $p$-ary cyclic codes,'' Des. Codes Cryptogr. doi: 10.1007/s10623-013-9908-2.

\bibitem{zhou}Z. Zhou, C. Ding, ``A class of three-weight cyclic codes,'' \emph{Finite Fields Appl.} \textbf{25}, 79-93(2014).
\bibitem{zhouIT} Z. Zhou,  C. Ding, J. Luo, A. Zhang, ``A family of five-weight cyclic codes and their weight enumerators,''
\emph{IEEE Trans. Inform. Theory}, \textbf{59}(10), 6674-6682(2013).


\end{thebibliography}
\end{document}